\documentclass[preprint,10pt,authoryear]{elsarticle}

\usepackage{graphicx, verbatim}

\journal{Information Processing \& Management}

\begin{document}

\begin{frontmatter}

\title{The skewness of computer science}

\author{Massimo Franceschet}

\address{Department of Mathematics and Computer Science, University of Udine \\
           Via delle Scienze 206 -- 33100 Udine, Italy \\
           \texttt{massimo.franceschet@dimi.uniud.it}}

\begin{abstract}
Computer science is a relatively young discipline combining science, engineering, and mathematics. The main flavors of computer science research involve the theoretical development of conceptual models for the different aspects of computing and the more applicative building of software artifacts and assessment of their properties. In the computer science publication culture, conferences are an important vehicle to quickly move ideas, and journals often publish deeper versions of papers already presented at conferences. These peculiarities of the discipline make computer science an original research field within the sciences, and, therefore, the assessment of classical bibliometric laws is particularly important for this field. In this paper, we study the skewness of the distribution of citations to papers published in computer science publication venues (journals and conferences). We find that the skewness in the distribution of mean citedness of different venues combines with the asymmetry in citedness of articles in each venue, resulting in a highly asymmetric citation distribution with a power law tail. Furthermore, the skewness of conference publications is more pronounced than the asymmetry of journal papers. Finally, the impact of journal papers, as measured with bibliometric indicators, largely dominates that of proceeding papers.
\end{abstract}


\end{frontmatter}

\section{Introduction} \label{introduction}

Computer science is an original discipline combining engineering and natural sciences as well as mathematics. It concerns itself with the representation and processing of information using algorithmic techniques. Research in computer science includes two main flavors: \textit{Theory}, developing conceptual frameworks for understanding the many aspects of computing, and \textit{Systems}, building software artifacts and assessing their properties \citep{CLMS09}.

A distinctive feature of computer science publication is the importance of prestigious conferences. Acceptance rates at selective computer science conferences range between 10\% and 20\%; for instance, in 2007-2008, ICSE (software engineering) 13\%, OOPSLA (object technology) 19\%, POPL (programming languages) 18\%. Journals have their role, but do not necessarily carry more prestige. The story of the development of computer science conferences is well reported by \citet{Fo09}, page 33:

\begin{quote}
\textit{The growth of computers in the 1950s led nearly every major university to develop a strong computer science discipline over the next few decades. As a new field, computer science was free to experiment with novel approaches to publication not hampered by long traditions in more established scientific and engineering communities. Computer science came of age in the jet age where the time spent traveling to a conference no longer dominated the time spent at the conference itself. The quick development of this new field required rapid review and distribution of results. So the conference system quickly developed, serving the multiple purposes of the distribution of papers through proceedings, presentations, a stamp of approval, and bringing the community together.}
\end{quote}

These peculiarities of the field -- the dualities Theory/System and Journal/Conference -- make computer science an original discipline within the sciences \citep{D05}. It is, therefore, interesting to investigate how these distinctive features of the discipline impact on the classical laws of informetrics \citep{B90}, e.g., Lotka's Law of scientific productivity \citep{L26}, Bradford's Law of scatter \citep{B34}, and skewness of citations to scientific publications \citep{S92}. In the present contribution, we study the skewness of the citation distribution of computer science papers. We distinguish between journal and conference papers, exploiting the  Conference Proceeding index recently added by Thomson Reuters to Web of Science. We furthermore tackle the problem of finding a theoretical model that well fits the empirical citation distributions. Finally, we compare the strength of impact of journal and proceeding papers as measured with bibliometric indicators, including the highly celebrated Hirsch index \citep{H05}.

The outline of the paper is as follows. In Section \ref{skewness} we study the shape of the citation distributions of computer science journal and conference papers. Section \ref{model} is devoted to the finding of a theoretical model that well fits such citation distributions. Finally, in Section \ref{conclusion} we summarize our findings and their implications.

\section{The skewness of citation distributions} \label{skewness}

Is the distribution of citations to computer science (CS) papers symmetric or skewed? A distribution is symmetric if the values are equally distributed around a typical figure (the mean);  a well-known example is the normal (Gaussian) distribution. A distribution is right-skewed if it contains many low values and a relatively few high values. It is left-skewed if it comprises  many high values and a relatively few low values. The power law distribution, for instance, is right-skewed. As a rule of thumb, when the mean is larger than the median the distribution is right-skewed and when the median dominates the mean the distribution is left-skewed. A more precise numerical indicator of distribution skewness is the third standardized central moment, that is the expected value of $(X - \mu)^3$, divided by $\sigma^3$, where $\mu$ and $\sigma$ are the mean and the standard deviation of the distribution of the random variable $X$, respectively. A value close to 0 indicates symmetry; a value greater than 0 corresponds to right skewness, and a value lower than 0 means left skewness.

In order to answer the posed research question about the skewness of the citation distribution of CS papers, we analysed the citations received by both CS journal and proceeding papers. As to journal articles, we accessed Thomson Reuters Journal Citation Reports (JCR), 2007 Science edition. The data source contains 281 computer science journals classified into the following six sub-fields corresponding to as many JCR subject categories: Artificial Intelligence (accounting for 29.2\% of the journals), Theory and Methods (27.8\%), Software Engineering (27.6\%), Information Systems (22.9\%), Hardware and Architectures (19.3\%), Cybernetics (7.2\%). Notice that the classification is overlapping. We purposely excluded category Interdisciplinary Applications, since the journals therein are only loosely related to computer science. For each journal title, we retrieved from Thomson Reuters Web of Science database all articles  published in the journal in 1999 (9,140 items in total) and the citations they received until 1st August 2009 (an overall of 106,849 references). A citation window of 9 years has been chosen because it corresponds to the mean cited half-life of CS journals tracked in Web of Science. The cited half-life for a journal is the median age of its papers cited in the current year. Half of citations to the journal are to papers published within the cited half-life. It follows that the citation state of the analysed papers is steady and will not \textit{significantly} change in the future.

As to conference papers, we used the Conference Proceedings Index recently added by Thomson Reuters to Web of Science database. Unfortunately, for this index, corresponding Proceedings Citation Reports are not yet published by Thomson Reuters. Furthermore, an annoying limitation of Web of Science is the impossibility of retrieving all papers belonging to a specific subject category. Therefore, we retrieved conference papers by country of affiliation addresses for authors. We took advantage of the country premier league compiled by \citet{K04}. The ranking contains countries in declining order with respect to the share of top 1\% of highly cited publications. Publications refer to period 1997-2001 and citations are collected in year 2002. The top-10 compilation reads: United States, United Kingdom, Germany, Japan, France, Canada, Italy, Switzerland, The Netherlands, and Australia. We hence retrieved all conference papers with at least one author affiliated to one of the mentioned top-10 countries that were published in 1999 and we tallied the citations they received until 1st August 2009. This amounts to 9,013 papers and 38,837 citations.

Journal papers are cited on average 11.69 times, the median (2nd quartile) is 4, the 1st quartile is 1 and the 3rd quartile is 11. The most cited journal paper received 1014 citations. The standard deviation, which measures the attitude of the data to deviate from the mean, is 31.66, that is, 2.71 times the mean. The average number of authors per paper is 2.33. The Hirsch index, or simply h index, is a recent bibliometric indicator proposed by \citet{H05}. The index, which found immediate interest both in the public \citep{Ba07} and in the bibliometrics community (see \citet{BD07} for opportunities and limitations of the measure), favors publication sets containing a continuous stream of influential works over those including many quickly forgotten ones or a few blockbusters. It is defined, for a publication set, as the highest number $n$ such that there are $n$ papers in the set each of them received at least $n$ citations. Geometrically, it corresponds to the size of the Durfee square contained in the Ferrers diagram of the citation distribution \citep{AHK08}. The Egghe index, or simply g index, is a variant of the h index measuring the highest number $n$ of papers that received together at least $n^2$ citations. It has been proposed by \citet{E06} to overcome some  limitations of the h index, in particular the fact that it disadvantages small but highly-cited paper sets too strongly.  We computed both indexes for the publication set of computer science journal papers: the h index amounts to 106, while the g index is 170.

\begin{figure}
\begin{center}
\includegraphics[scale=0.35, angle=-90]{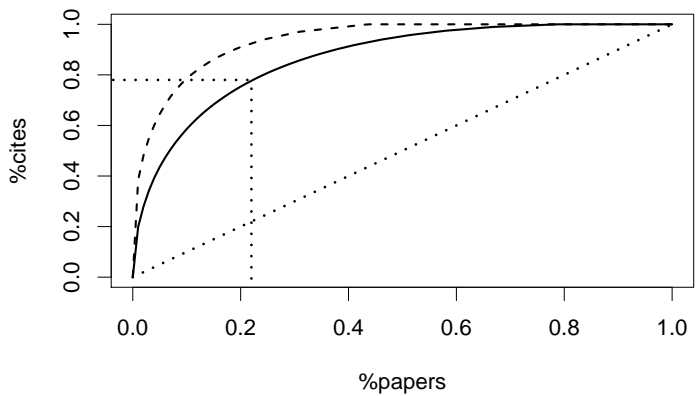}
\caption{Lorenz curve: share of most cited journal articles (solid curve) and conference articles (dashed curve) versus share of citations received by the articles. The dotted line with slope 1 corresponds to the hypothetical situation in which each article is equally cited. The vertical and horizontal dotted lines illustrate the Pareto principle for journal papers: 22\% of the most cited papers account for 78\% of the citations.}
\label{Pareto}
\end{center}
\end{figure}

The citation distribution is right skewed (Figure~\ref{Pareto} depicts the Lorenz curve): 76\% of the papers are cited less than the average and 21\% are uncited. The most cited 7\%  of the articles collect more than half of the citations; these papers are cited, on average, 13 times more than the other papers. The most cited half of the  articles harvest 95\% of the citations; these papers are cited, on average, 19 times more than the other papers. In particular, we noticed that 78\% of the citations come from 22\% of the papers, matching quite well the Pareto principle or 80-20 rule, which claims that 80\% of the effects come from 20\% of the causes \citep{P97}. The skewness indicator is 13.08, well beyond the symmetry value of 0. The Gini index is a measure of concentration of the character (citations) among the statistical units under consideration (journals). The two extreme situations are equidistribution, in which each journal  receives the same amount of citations (the Gini index is equal to 0), and maximum concentration, in which the total amount of the citations is attributed to a single journal (the Gini index is equal to 1). The Gini index for the journal citation distribution is 0.73, indicating a high concentration of citations among journal papers.

Proceedings papers are cited on average 4.31 times, significantly less than journal articles (the median is 0). Nevertheless, the most cited proceeding paper collects a whopping number of citations (2707). Standard deviation is 34.11, or 7.9 times the mean. Citations to conference papers deviate even more from the average citation value than do citations to journal articles. The conference h index is 65, which amounts to 61\% of the journal h index. The conference g index is 129, or 76\% of its journal counterpart. Notice that the disproportionately large number of citations harvested by the top-cited conference paper significantly shorten the gap between the conference g index and its journal counterpart, whereas this citational blockbuster has little influence on the conference h index score. The average number of authors per paper (2.85) is higher than that for journal papers, meaning that computer science authors are more motivated to collaborate when writing proceeding papers.

The distribution of citations to conference papers is even more skewed than that of journal papers: the concentration curve for conference articles markedly dominates that for journal papers  (Figure~\ref{Pareto}). Indeed, the majority of proceeding papers (56\%) sleep uncited, and 84\% of the papers are cited less than the average. The most cited 3\% of the papers harvest more than half of the citations, and 85\% of the citations come from 15\% of the papers. The skewness indicator amounts to 57.94, much higher than the value computed for journal papers. Citations are highly concentrated among few conference papers, as indicated by the Gini index, which amounts to 0.88, a higher score with respect to what we measured for journal articles.

We observed a certain degree of skewness in the distribution of citations to articles published in each venue (journal or conference) as well. For instance, the 135 papers published in 1999 in the flagship ACM magazine, \textit{Communications of the ACM}, received 3003 citations, with an average of 22 citations per paper, which largely dominates the median citation rate (8) and is even greater than the  third quartile (20). A share of 76\% of the papers are cited less than the average paper. The distribution skewness is 3.65 and the coefficient of variation is 1.82. The flagship IEEE magazine, \textit{IEEE Computer}, published in the same year 130 papers receiving 1346 citations, a mean of about 10 citations per paper, which is bigger than the median value (2) and close to the third quartile (11). The percentage of the papers cited less than the average is 73\%. The distribution skewness is 4.01 and the coefficient of variation amounts to 2.09.

Two related conferences are \textit{ACM SIGMOD International Conference on Management of Data} (held in Philadelphia, Pennsylvania, in 1999) and  \textit{IEEE International Conference on Data Engineering} (located in Sydney, Australia, in 1999). SIGMOD published 76 papers with an average of 3 citations per paper, a median of 1 citation, and a maximum of 26 citations. Three papers over four are cited less than the average article; the distribution skewness is 2.54 and the  coefficient of variation amounts to 1.71. ICDE published 67 papers; the average paper collects 10 citations and the median one has 3 citations. The top-cited article received 102 citations. A share of 81\% of the published articles receive less citations than the average article. Skewness is at 3.23 and variation amounts to 1.94.

Interestingly, the skewness of citations to papers within venues, albeit significant, is less noticeable than the asymmetry of citations to all papers in the field. This might indicate that venues (journals or conferences) represent citational homogeneous samples of the field.

The distribution of mean citedness of different journals is also skewed. The $n$-year impact factor for a journal with respect to a fixed census year and a target window of $n$ years is the mean number of citations that occurred in the census year to the articles published in that journal during the previous $n$ years \citep{Ga06}. Typical target windows are 2 and 5 years long. We analysed the distribution of 2007 impact factors of CS journals. The mean 2-year impact factor is 1.004 which is greater than the median 2-year impact factor that is equal to 0.799; the distribution skewness is 1.64. The mean 5-year impact factor is 1.218 and dominates the median 5-year impact factor that is equal to 0.914; the distribution skewness is 2.69. Again, the skewness of journal mean citedness is less important than the asymmetry of citations to all papers in the field. Unfortunately, till today, Thomson Reuters does not provide impact factor scores for conference proceedings.

\section{A theoretical model for the citation distributions} \label{model}

What theoretical model best fits the empirical citation distribution for CS papers? Having a theoretical model that well fits the citation distribution would increase our understanding of the dynamics of the underlying citational complex system. We compared our empirical samples with the following three well-known right-skewed distributions.

The \textit{power law distribution}, also known as \textit{Pareto distribution},  is named after the Italian economist Vilfredo Pareto who originally observed it studying the allocation of wealth among individuals: a larger share of wealth of any society (approximately 80\%) is owned by a smaller fraction (about 20\%) of the people in the society \citep{P97}. Examples of phenomena that are Pareto distributed include: degree of interaction (number of distinct interaction partners)  of proteins, degree of nodes of the Internet, intensity (number of battle deaths) of wars, severity (number of deaths) of terrorist attacks, number of customers affected in electrical blackouts, number of sold copies of bestselling books, size of human settlements, intensity of solar flares, number of religious followers, and frequency of occurrence of family names (see \citet{CSN09} and references therein). Bibliometric phenomena that provably follow a power law model are word frequency in relatively lengthy texts \citep{Z49,CSN09}, scientific productivity of scholars \citep{L26,CSN09}, and, interestingly, number of citations received between publication and 1997 by top-cited scientific papers published in 1981 in journals catalogued by the ISI (the former name of Thomson Reuters) \citep{R98}. The probability density function for a Pareto distribution is defined for $x \geq x_0 > 0$ in terms of the scaling exponent parameter $\alpha > 1$ as follows:
$$
f(x) = \frac{(\alpha-1) \, x_{0}^{\alpha-1}}{x^{\alpha}}
$$

The \textit{stretched exponential distribution} is a family of extensions of the well-known exponential distribution characterized by fatter tails. \citet{LS98} show that different phenomena in nature and economy can be described in the regime of the exponential distribution, including radio and light emission from galaxies, oilfield reserve size, agglomeration size, stock market price variation, biological extinction event, earthquake size, temperature variation of the earth, and citation of the most cited physicists in the world. The probability density function is a simple extension of the exponential distribution with one additional stretching parameter $\alpha$:
$$
f(x) = \alpha \lambda^\alpha x^{\alpha -1} e^{-(\lambda x)^\alpha}
$$
where $x \geq 0$, $\lambda > 0$ and $0 < \alpha \leq 1$. In particular, if the stretching parameter $\alpha = 1$, then the distribution is the usual exponential distribution. When the parameter $\alpha$ is not bounded from $1$, the resulting distribution is better known as the \textit{Weibull distribution}.

The \textit{lognormal distribution} is the distribution of any random variable whose logarithm is normally distributed. Phenomena determined by the multiplicative product of many independent effects are characterized by a lognormal model. The lognormal distribution is a usual suspect in bibliometrics. In a study based on the publication record of the scientific research staff at Brookhaven National Laboratory, \citet{S57} observes that the scientific publication rate is approximately lognormally distributed. More recently, \citet{SSN08} study the citation distribution for individual journals indexed in Web of Science and show that there exists a steady state period of time, specific to each journal, such that the number of citations to papers published in the journal in that period will not significantly change in the future. They also demonstrate that, with respect to the journal steady state period, the citations to papers published in individual journals follow a lognormal model. Finally, \citet{RFC08} analyse the distribution of the ratio between the number of citations received by an article and the average number of citations received by articles published in the same field and year for papers in different sub-fields corresponding to Thomson Reuters JCR categories (the category closest to computer science is cybernetics). They find a similar distribution for each category with a good fit with the lognormal distribution.

The lognormal probability density function is defined in terms of parameters $\mu$ and $\sigma > 0$ as follows:
$$
f(x) = \frac{1}{x \sigma \sqrt{2 \pi}} e^{-\frac{(\log(x)-\mu)^2}{2 \sigma^2}}
$$
for $x > 0$.

We compared the empirical article citation distributions of journal and conference articles and the mentioned theoretical models with the following methodology \citep{CSN09}:

\begin{enumerate}
\item we gauge the distribution parameters using the \textit{maximum likelihood estimation method} (MLE), which finds the parameters that maximize the likelihood of the data with respect to the model;

\item we estimate the goodness-of-fit between the empirical data and a theoretical model taking advantage of the Kolmogorov-Smirnov (KS) test. The test compares an empirical and a theoretical model by computing the maximum absolute difference between the empirical and theoretical cumulative frequencies (this distance is the KS statistic). To appreciate if the measured distance is statistically significant\footnote{Unfortunately, the fitness significance (p-value) computed by the Kolmogorov-Smirnov test is known to be biased if the parameters of the theoretical model are not fixed but, instead, they are estimated from the observed data.}, we adopted the following Monte Carlo procedure, as suggested in \citet{CSN09}:

\begin{enumerate}
\item we compute the KS statistic for the empirical data and the theoretical model with the MLE parameters estimated for the empirical data;

\item we generate a large number of synthetic data sets following the theoretical model with the MLE parameters estimated for the empirical data;

\item for each synthetic data set, we compute its own MLE parameters and fit it to the theoretical model with the estimated parameters (and not to the model with the parameters of the original distribution from which the data set is drawn). We record the KS statistic for the fit;

\item we count what fraction of the time the resulting KS statistic for synthetic data is larger than or equal to the KS statistic for the empirical data. This fraction measures the fitness significance (\textit{p-value}).
\end{enumerate}
\end{enumerate}

Following \citet{CSN09}, we generated 2500 synthetic data sets. This guarantees that the p-valued is accurate to 2 decimal digits. Moreover, the hypothesis of goodness of fit of the observed data with respect to the theoretical model is ruled out if the p-value is lower than 0.1, that is, if less than 10\% of the time the distance of the observed data from the model is dominated by the very same distance for synthetic data. We performed all statistical computations using R \citep{R08}.

Table \ref{KS} contains the results of our tests. For both journal and conference data sets, the best fit is achieved by the lognormal model. Furthermore, for each surveyed theoretical model, the journal citation distribution fits better the model than the conference counterpart. Nevertheless, the computed p-values are not statistically significant, hence we cannot accept the hypothesis that the \textit{entire} observed citation distributions follow one of the surveyed theoretical distributions.

\begin{table}
\begin{center}
\begin{tabular}{|l|c|c|c|c|c|c|c|c|c|}
\hline
\textbf{Data set} &
\multicolumn{3}{c}{\textbf{Pareto}} &	
\multicolumn{3}{|c|}{\textbf{Stretched Exp}} &	
\multicolumn{3}{|c|}{\textbf{Lognormal}} \\ \hline
 &
 $\alpha$ & $x_0$ & \textbf{KS} &
 $\alpha$ & $\lambda$ & \textbf{KS} &
 $\mu$    & $\sigma$  & \textbf{KS}  \\ \hline
\textbf{Journal}    & 1.55 & 1 & 0.19 & 0.75 & 0.08 & 0.14  & 1.84 & 1.25 & 0.08      \\ \hline
\textbf{Conference} & 1.78 & 1 & 0.29 & 0.70 & 0.15 & 0.23  & 1.29 & 1.18 & 0.15   \\ \hline
\end{tabular}
\end{center}
\caption{MLE distribution parameters and Kolmogorov-Smirnov statistic. All p-values are not significant.}
\label{KS}
\end{table}

In practice, few empirical phenomena obey power laws on the entire domain. More often the power law applies only for values greater than or equal to some minimum $x_0$. In such case, we say that the \textit{tail} of the distribution follows a power law. For instance, \citet{CSN09} analysed 24 real-world data sets from a range of different disciplines, each of which has been conjectured to follow a power law distribution in previous studies. Only 17 of them passed the test with a p-value of at least 0.1, and all of them show the best adherence to the model when a suffix of the distribution is considered.  Notable phenomena that were ruled out from the Pareto fitting are  size of files transmitted on the Internet, number of hits to web pages, and number of links to web sites. For the latter two, the lognormal model represents a more plausible hypothesis. The relative sizes of the tails with respect to the size of the entire distribution for the power law distributed phenomena range from 0.002 to 0.61, with a median relative tail length of 0.11. In particular, for the data set containing citations received by scientific papers in journals catalogued by the ISI \citep{R98}, the relative tail size  is 0.008, meaning that only the distribution of citations to articles cited at least 160 times well fits the Pareto model.

We tested the hypothesis that a significantly large tail of the distribution of citations to computer science papers follows a power law model. For the estimation of the lower bound $x_0$ parameter, that is, the starting value of the distribution tail, we followed the approach put forward by \citet{CYG07}. The idea behind this method is to choose the value of $x_0$ that makes the empirical probability distribution and the best-fit power law model as similar as possible beginning from $x_0$. We used the KS statistic to gauge the distance between the observed data and the theoretical ones. Finally, we estimate the significance of the goodness-of-fit between the empirical data and the best-fit power law model following the above-described Monte Carlo procedure.

\begin{figure}
\begin{center}
\includegraphics[scale=0.35, angle=-90]{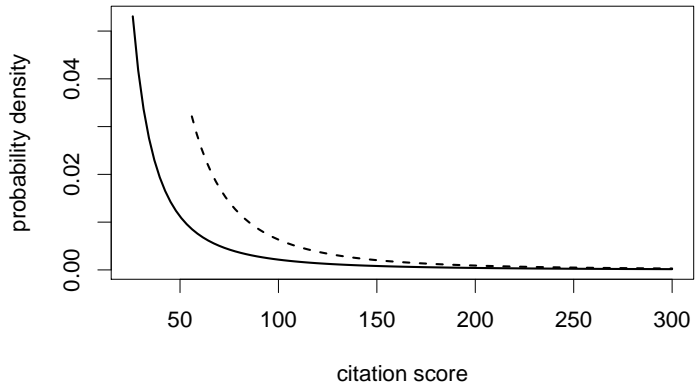}
\caption{Probability density functions for journal articles (dashed curve) and conference papers (solid curve) starting from the corresponding lower cutoffs.}
\label{powerlaw}
\end{center}
\end{figure}

The results are that the citations to articles published in computer science journals that are cited at least 56 times indeed follow a power law distribution with scaling exponent $\alpha = 2.80$. This means that the probability density distribution for citations to journal articles is $C / x^{2.80}$, where $C = 2525$ and $x \geq x_0 = 56$.  The KS statistic is 0.033 and the computed p-value is 0.38, well beyond the significance threshold of 0.1. The Pareto-distributed tail is 355 articles long, or 4\% of the entire distribution. As to conference papers, the power law behaviour shows up for articles cited at least 26 times, which corresponds to a distribution tail of 260 articles, or 3\% of the entire data set.  The scaling exponent  $\alpha = 2.38$ and, hence, the probability density distribution reads $C / x^{2.38}$, where $C = 124$ and $x \geq x_0 = 26$. The KS statistic is 0.046 and the computed p-value is 0.23; the fit is hence statistically significant but less good than what computed for journal papers. Figure \ref{powerlaw} depicts the theoretical models for journal and conference papers starting from the corresponding lower thresholds. Notice that the conference scaling exponent (2.38) is lower than the journal exponent (2.80), meaning that the asymptotic decay of the probability density function is slower for conference papers than for journal ones. The journal multiplicative constant (2525) is, however, much bigger than the conference counterpart (124). The consequence is that the probability density for journal papers dominates the probability density for conference papers up to a certain citation value, showing, up to this point, a fatter tail as depicted in Figure \ref{powerlaw}. From that point onwards, however, the asymptotic behaviour shows up, and the conference tail results heavier, a consequence of the  extraordinary number of citations (2707) collected by the top-cited conference paper. The meeting point of the two curves is, however, around 1313, above the biggest citation score for journal papers (1014). 

It is worth mentioning that, according to our experiments, the \textit{best} cutoff, that is, the starting point that minimizes the KS statistic, is also the the lowest \textit{good enough} cutoff, that is the smallest threshold that guarantees a power law fitting with a p-value of at least 0.1. In other words, the cutoff we have found guarantees both that the distance from the theoretical model is minimum and that the length of the distribution tail starting at the cutoff is maximum. Finally, we checked that both the lognormal and the stretched exponential models do not fit well the identified power-law distributed tails.

\section{Discussion} \label{conclusion}

Our main findings and their implications are summarized in the following.

\medskip
\noindent
\textit{The citation distribution for computer science papers is severely skewed.}
Such an extreme asymmetry is the combination of the skewness of mean citedness of different venues and of the skewness of citedness of articles published in each venue. A similar two-leveled citational hierarchy has been noticed by \citet{S92} in the field of biomedicine when authors (and not journals) are taken to represent functional units of the scientific system.

Skewness of citation distribution has an important consequence: the mean is not the appropriate measure of the central tendency of the citations received by articles.
Indeed, only a small number of articles are cited near or above the mean value and the great majority of them are endorsed less than the average, with a significant share of the papers that sleep uncited. Assigning the same value to all articles levels out the differences that evaluation procedures should highlight \citep{S92}. A more appropriate measure of central tendency in case of skewed distributions is the \textit{median} (see also \citet{W09} and \citet{CB09}).

Since the Thomson Reuters impact factor is, roughly, the mean number of recent citations received by papers published in a given journal, such a popular measure of journal impact is not immune to the skewness property of citation distributions and, therefore, it might be misused. While sorting journals according to the mean or the median can yield rankings that statistically differ little overall, the absolute magnitude in the differences in mean citedness between journals is oftentimes misleading \citep{W09}. Similarly, it is biased to gauge the impact of individual papers or authors using the impact factor of the publishing journals \citep{Ga06,P09}. It would be more fair to judge individual contributions and their contributors by their own citation scores, as soon as this data is robustly available.

A simple example might better convince the reader. Let A and B be journals such that each of them published 4 papers. Suppose papers in journal A are cited 1, 1, 1, and 97 times, respectively, while those in journal B are each cited 25 times. Clearly, the average paper in both journals has the same number of citations (4). However, can we conclude that journals A and B have the same impact score? To find a good answer, we have to start from the right question. My assessment is that the appropriate way to pose the question is: what is the  \textit{probability} that for two randomly drawn papers $P$ and $Q$ published in journals $A$ and $B$, respectively, the number of citations of $P$ is greater than the number of citations of $Q$? This probability is, interestingly, rather low: 25\% (the top-cited A paper beats all B papers, but any other A paper loses against any B paper). Hence, while the mean citedness assigns the same impact to both journals, there are high changes to find better papers in journal B. Put another way, I would certainly buy journal B, if I were a librarian facing the choice to purchase only one of the two journals due to budget limitations.

The reader might rightly argue that this is an artificial example, with scant bearing on real journal citation records. Indeed, the citation record of journal B is rather uncommon. Hence, let us consider a real example. IEEE Transactions on Information Theory (TIT) published 364 articles during period 1997-1999 that received an average of 26 citations until 1st August 2009. IEEE Transactions on Computers (TC) issued 375 articles in the same period, collecting an average of 13 citations. Hence, the mean impact of TIT is twice as big as the mean impact of TC. Nevertheless, both journals have a median impact of 8 citations. Even more amazingly, the probability of finding a higher cited paper in TIT is only 50.7\%, the probability of finding a higher cited paper in TC is 45.5\%, while in 3.8\% of the cases we have a tie. The surprise vanishes as soon as we compute the relative deviation from the mean (coefficient of variation) for the two journal distributions: this is significantly larger for TIT (2.68) than for TC (1.34). Curiously, the deviation of TIT is exactly twice that of TC.

\medskip
\noindent
\textit{The tail of the citation distribution for computer science papers has a power law behaviour.}

Networks with power law node degree distribution have been extensively studied: \citet{B03} provides a captivating and elegantly written introduction to the field. These graphs are referred to as \textit{scale-free networks}: they show a continuous hierarchy of nodes, spanning from rare kings to the numerous tiny nodes, with no single node that might be considered to be characteristic of all the nodes.  By contrast, in random networks the degree distribution resembles a bell curve, with the peak of the distribution corresponding to the characteristic scale of node connectivity. In a random citation network, the vast majority of articles receive the same number of citations, and both poorly and highly endorsed papers are extremely rare.

Articles with a truly extraordinary knack of grabbing citations are the \textit{authorities} of the citation network. Articles, like review papers, that cite a considerably number of references are the network \textit{hubs} \citep{K99}. Highly-cited review papers are both authorities and hubs: they are \textit{connectors}, with the peculiar ability to relate ostensibly different topics and to create short citation paths between any two nodes in the system, making the citation network look like a \textit{small world}.

The emergence of scale-free networks has been theoretically explained with a simple model encompassing \textit{growth} and \textit{preferential attachment}  \citep{BA99,BAJ99}. According to this model, the network starts from a small nucleus and expands with the addition of new nodes. The new nodes, when deciding where to link, prefer to attach to the nodes having more links. Preferential attachment matches the previously investigated bibliometric principle of \textit{cumulative advantage}: a paper which has been cited many times is more likely to be cited again than one which has been little cited \citep{P76}. Moreover, extraordinary citation scores may be also the consequence of a number of recognized citation biases, including advertising (self-citations), comradeship (in-house citations), chauvinism, mentoring, obliteration by incorporation, flattery, convention, and reference copying \citep{MM89}.

\medskip
\noindent
\textit{The impact of journal articles, as gauged with the aid of bibliometrics, is significantly higher than the impact of conference papers.}

The role of conference publications in computer science is controversial. Conferences provide fast and regular publication of papers, which is particularly important since computer science is a relatively young and fast evolving discipline. Moreover, conferences help to bring researchers together. It is not a mere coincidence that the average conference article has more authors than the typical journal paper.  Lately, however, many computer scientists highlighted  many flaws of the conference systems, in particular when compared to archival journals \citep{R09,BS09,V09,Fo09}. \citet{F10-CACM}  gives a bibliometric perspective on the role of conferences in computer science and concludes that, wearing bibliometric lens, the best strategy to gain impact is that of publishing few, final, and well-polished contributions in archival journals, instead of many premature `publishing quarks' in conference proceedings. The present contribution reinforces these conclusions.


\begin{thebibliography}{37}
\expandafter\ifx\csname natexlab\endcsname\relax\def\natexlab#1{#1}\fi
\expandafter\ifx\csname url\endcsname\relax
  \def\url#1{\texttt{#1}}\fi
\expandafter\ifx\csname urlprefix\endcsname\relax\def\urlprefix{URL }\fi

\bibitem[{Anderson et~al.(2008)Anderson, Hankin, and Killworth}]{AHK08}
Anderson, T.~R., Hankin, R. K.~S., Killworth, P.~D., 2008. Beyond the {Durfee}
  square: enhancing the h-index to score total publication output.
  Scientometrics 76~(3), 577--588.

\bibitem[{Ball(2007)}]{Ba07}
Ball, P., 2007. Achievement index climbs the ranks. Nature 448, 737.

\bibitem[{Barab\'{a}si(2003)}]{B03}
Barab\'{a}si, A.-L., 2003. Linked. Perseus Publishing.

\bibitem[{Barab\'{a}si and Albert(1999)}]{BA99}
Barab\'{a}si, A.-L., Albert, R., 1999. Emergence of scaling in random networks.
  Science 286, 509--512.

\bibitem[{Barab\'{a}si et~al.(1999)Barab\'{a}si, Albert, and Jeong}]{BAJ99}
Barab\'{a}si, A.-L., Albert, R., Jeong, H., 1999. Mean-field theory for
  scale-free random networks. Physica A 272, 173--187.

\bibitem[{Birman and Schneider(2009)}]{BS09}
Birman, K., Schneider, F.~B., 2009. Program committee overload in systems.
  Communications of the ACM 52~(5), 34--37.

\bibitem[{Bookstein(1990)}]{B90}
Bookstein, A., 1990. Informetric distributions, part {I}: Unified overview.
  Journal of the American Society for Information Science 41, 368--375.

\bibitem[{Bornmann and Daniel(2007)}]{BD07}
Bornmann, L., Daniel, H.-D., 2007. What do we know about the h index? Journal
  of the American Society for Information Science and Technology 58~(9),
  1381--1385.

\bibitem[{Bradford(1934)}]{B34}
Bradford, S.~C., 1934. Sources of information on specific subjects. Engineering
  137, 85--86.

\bibitem[{Calver and Bradley(2009)}]{CB09}
Calver, M.~C., Bradley, J.~S., 2009. Should we use the mean citations per paper
  to summarise a journal's impact or to rank journals in the same field?
  Scientometrics 81~(3), 611--615.

\bibitem[{Choppy et~al.(2009)Choppy, van Leeuwen, Meyer, and
  Staunstrup}]{CLMS09}
Choppy, C., van Leeuwen, J., Meyer, B., Staunstrup, J., 2009. Research
  evaluation for computer science. Communications of the {ACM} 52~(4), 31--34.

\bibitem[{Clauset et~al.(2009)Clauset, Shalizi, and Newman}]{CSN09}
Clauset, A., Shalizi, C.~R., Newman, M. E.~J., 2009. Power-law distributions in
  empirical data. SIAM Review 51, 661--703.

\bibitem[{Clauset et~al.(2007)Clauset, Young, and Gleditsch}]{CYG07}
Clauset, A., Young, M., Gleditsch, K.~S., 2007. On the frequency of severe
  terrorist events. Journal of Conflict Resolution 51~(1), 58--87.

\bibitem[{de~Solla~Price(1976)}]{P76}
de~Solla~Price, D., 1976. A general theory of bibliometric and other cumulative
  advantage processes. Journal of the American Society for Information Science
  27, 292--306.

\bibitem[{Denning(2005)}]{D05}
Denning, P.~J., 2005. Is {Computer Science} science? Communications of the
  {ACM} 48~(5), 27--31.

\bibitem[{Egghe(2006)}]{E06}
Egghe, L., 2006. Theory and practice of the g-index. Scientometrics 69~(1),
  131--152.

\bibitem[{Fortnow(2009)}]{Fo09}
Fortnow, L., 2009. Time for computer science to grow up. Communications of the
  ACM 52~(8), 33--35.

\bibitem[{Franceschet(2010)}]{F10-CACM}
Franceschet, M., 2010. The role of conference publications in computer science:
  a bibliometric view. Communications of the {ACM}. In press.

\bibitem[{Garfield(2006)}]{Ga06}
Garfield, E., 2006. The history and meaning of the journal impact factor.
  Journal of the American Medical Association 295~(1), 90--93.

\bibitem[{Hirsch(2005)}]{H05}
Hirsch, J.~E., 2005. An index to quantify an individual's scientific research
  output. Proceedings of the National Academy of Sciences of USA 102~(46),
  16569--16572.

\bibitem[{King(2004)}]{K04}
King, D.~A., 2004. The scientific impact of nations. Nature 430, 311--316.

\bibitem[{Kleinberg(1999)}]{K99}
Kleinberg, J.~M., 1999. Authoritative sources in a hyperlinked environment.
  Journal of the ACM 46~(5), 604--632.

\bibitem[{Laherr\'{e}re and Sornette(1998)}]{LS98}
Laherr\'{e}re, J., Sornette, D., 1998. Stretched exponential distributions in
  nature and economy: ``fat'' tails with characteristic scales. The European
  Physical Journal B 2, 525--539.

\bibitem[{Lotka(1926)}]{L26}
Lotka, A.~J., 1926. The frequency distribution of scientific productivity.
  Journal of the Washington Academy of Sciences 16, 317--323.

\bibitem[{MacRoberts and MacRoberts(1989)}]{MM89}
MacRoberts, M.~H., MacRoberts, B.~R., 1989. Problems of citation analysis: A
  critical review. Journal of the American Society for Information Science
  40~(5), 342--349.

\bibitem[{Pareto(1897)}]{P97}
Pareto, V., 1897. Cours d'\'{e}conomie politique. Vol.~2. Universit\'{e} de
  Lausanne, Lausanne.

\bibitem[{Pendlebury(2009)}]{P09}
Pendlebury, D.~A., 2009. The use and misuse of journal metrics and other
  citation indicators. Archivum Immunologiae et Therapiae Experimentalis
  57~(1), 1--11.

\bibitem[{{R Development Core Team}(2008)}]{R08}
{R Development Core Team}, 2008. R: A Language and Environment for Statistical
  Computing. R Foundation for Statistical Computing, Vienna, Austria, {ISBN}
  3-900051-07-0. \texttt{http://www.R-project.org}.

\bibitem[{Radicchi et~al.(2008)Radicchi, Fortunato, and Castellano}]{RFC08}
Radicchi, F., Fortunato, S., Castellano, C., 2008. Universality of citation
  distributions: Toward an objective measure of scientific impact. Proceedings
  of the National Academy of Sciences of USA 105~(45), 17268--17272.

\bibitem[{Redner(1998)}]{R98}
Redner, S., 1998. How popular is your paper? {An} empirical study of the
  citation distribution. The European Physical Journal B 4, 131--134.

\bibitem[{Reed(2009)}]{R09}
Reed, D., 2009. Publishing quarks: Considering our culture. Computing Research
  News 21~(2).

\bibitem[{Seglen(1992)}]{S92}
Seglen, P.~O., 1992. The skewness of science. Journal of the American Society
  for Information Science 43~(9), 628--638.

\bibitem[{Shockley(1957)}]{S57}
Shockley, W., 1957. On the statistics of individual variations of productivity
  in research laboratories. Proceedings of the IRE 45, 279--290.

\bibitem[{Stringer et~al.(2008)Stringer, Sales-Pardo, and Amaral}]{SSN08}
Stringer, M.~J., Sales-Pardo, M., Amaral, L. A.~N., 2008. Effectiveness of
  journal ranking schemes as a tool for locating information. PLoS ONE 3~(2),
  e1683.

\bibitem[{Vardi(2009)}]{V09}
Vardi, M., 2009. Conferences vs.\ journals in computing research.
  Communications of the ACM 52~(5), 5.

\bibitem[{Wall(2009)}]{W09}
Wall, H.~J., 2009. Don't get skewed over by journal rankings. The {B.E.}
  Journal of Economic Analysis \& Policy 9~(1), 1--10.

\bibitem[{Zipf(1949)}]{Z49}
Zipf, G.~K., 1949. Human behavior and the principle of least effort.
  Addison-Wesley.

\end{thebibliography}
\end{document}